\documentstyle[prl,aps,epsf,twocolumn]{revtex}

\begin{document}
\draft
\title{Implications of non-feasible transformations among icosahedral 
$\bbox{h}$ orbitals}
\author{Edwin Lo and B.~R.~Judd}
\address{Henry A Rowland Department of Physics and Astronomy, 
The Johns Hopkins University, Baltimore, Maryland 21218}
\date{\today}
\maketitle
\begin{abstract}
The symmetric group $S_6$ that permutes the six five-fold axes of an
icosahedron is introduced to go beyond the simple rotations that constitute
the icosahedral group $I$. Owing to the correspondence $h\leftrightarrow 
{\rm d}$, the calculation of the Coulomb energies for the icosahedral 
configurations $h^N$ based on the sequence ${\rm O(5)} \supset S_6 \supset 
S_5 \supset I$ can be brought to bear on Racah's classic theory for the 
atomic d shell based on ${\rm SO(5)} \supset {\rm SO}_L(3) \supset I$. Among 
the elements of $S_6$ is the kaleidoscope operator ${\cal K}$ that rotates the 
weight space of SO(5) by $\frac{\pi}2$. Its use explains some puzzling 
degeneracies in d$^3$ involving the spectroscopic terms $^2$P, $^2$F, $^2$G 
and $^2$H.
\end{abstract}

One of the most enduring mysteries in the theoretical
analysis of the atomic d shell was noticed over 60 years ago by Condon and
Shortley\cite{CS}.  They remarked that the non-relativistic Coulomb energies
for the terms $^2$P and $^2$H of d$^3$ are degenerate, but they did not ask 
why.  The degeneracy can be brought into focus by breaking the Coulomb
interaction into three parts $e_i \ (i=0,1,2)$, following the work of
Racah\cite{Racah}.  Both $e_0$ and $e_1$ are scalars with respect to SO(5),
the rotation group spanning the five orbital states of a d electron, and for
these two operators the degeneracy is expected.  However, $e_2$ belongs to
the irreducible representation (IR) (22) of SO(5), and there is no reason to
anticipate the theoretical result
\begin{equation}
\langle {\rm d^3\,^2P}\,|\,e_2\,|\,{\rm d^3\,^2P}\rangle =
\langle {\rm d^3\,^2H}\,|\,e_2\,|\,{\rm d^3\,^2H}\rangle \label{couldeg}
\end{equation}
which is moderately well observed in the spectra of transition-metal
ions\cite{Sugar}.  Advances in the analyses of such spectra have
revealed similar equalities over the years.  For example, interactions
between d$^N$ and other configurations can be represented by operators
acting within the d shell on more than two electrons at a time, and the
three-electron operator $t_3$ belonging to the IR (30) of SO(5) has been
found to satisfy the equations\cite{Leavitt}
\begin{eqnarray}
\langle {\rm d^3\,^2P}\,|\,t_3\,|\,{\rm d^3\,^2P}\rangle &=&\langle 
{\rm d^3\,^2H}\,|\,t_3\,|\,{\rm d^3\,^2H}\rangle \ , \label{t3deg1} \\
\langle {\rm d^3\,^2F}\,|\,t_3\,|\,{\rm d^3\,^2F}\rangle &=&\langle 
{\rm d^3\,^2G}\,|\,t_3\,|\,{\rm d^3\,^2G}\rangle \ , \label{t3deg2}
\end{eqnarray}
neither of which has received an explanation.

Work on the icosahedral molecule C$_{60}$ has revived interest in the atomic
d shell.  This has come about because, on restricting the rotation group
SO(3) in ordinary space to its icosahedral subgroup $I$, the five orbital
components of a d electron serve as a basis for the IR $H$ of $I$.  (This
IR is distinguished from the spectroscopic term $^2$H by italicizing it.)
The other IRs of $I$ are $A,T_1,T_2$ and $G$, of dimensions 1,3,3 and 4
respectively. Following convention, we use lower-case letters for individual
electron states.  Last year, Oliva\cite{Oliva} took advantage of the
correspondence $h^N \leftrightarrow {\rm d}^N$ to study the Coulomb energies
of the icosahedral configuration $h^3$.  Instead of the three SO(3) scalars
$e_i$ of Racah, he found that there were five two-electron icosahedral
scalars, three of which coincide with Racah's operators.  In spite of the
existence now of five operators of various strengths ${\cal H}^k$ (called {\em 
molecular invariants} by Oliva), it turns out that, for any choice of ${\cal 
H}^k$, one of the two $^2T_1$ levels always coincides with a $^2T_2$ level.  
The eigenfunction of that $^2T_1$ level is a mixture of states that, in the 
atomic picture, correspond to $^2$P and $^2$H, and which therefore appear to 
be connected to Eqs.(\ref{couldeg}) and~(\ref{t3deg1}) above. Oliva 
speculated that an additional symmetry operation transforming $^2$H into 
$^2$P and vice versa should exist.  It is this kind of possibility that we 
wish to explore in the present Letter.

Because of the correspondence $h\leftrightarrow{\rm d}$, any problem for
$h^N$ can be recast for d$^N$.  The advantage of using d electrons lies in
being able to call upon the familiar theory for SO(3).  Each of the two new
Coulomb operators, for example, can be expressed in terms of the components
of a spherical tensor $\bbox{T}^{(6)}$ (this having the lowest nonzero rank 
that can provide an icosahedral scalar\cite{Harter}).  We write
\begin{equation}
e^{(6)} = \case{\sqrt{11}}5T^{(6)}_0+\case{\sqrt7}5\left( T^{(6)}_5-
T^{(6)}_{-5}\right)
\end{equation}
in which
\begin{equation}
\bbox{T}^{(6)}=\sum_{i\ne j}\left\{a(\bbox{v}^{(2)}_i\bbox{v}^{(4)}_j)^{(6)}
+b(\bbox{v}^{(4)}_i \bbox{v}^{(4)}_j)^{(6)}\right\} , \label{e23}
\end{equation}
where the reduced matrix element $({\rm d}\| v^{(k)}\| {\rm d})$ for electron
$i$ or $j$ is arbitrarily set at $\sqrt{2k+1}$.  The two operators
$e^{(6)}_2$ and $e^{(6)}_3$ are defined by taking $(a,b)=(4,-2\sqrt5)$ and 
$(10,2\sqrt5)$ respectively, and correspond to the
IRs (22) and (40) of SO(5).  Such operators represent the effect to
second-order in perturbation theory of a d-shell atom placed in an electric
field of icosahedral symmetry, there being no first-order effect because
an electric potential of rank 6 cannot split a d state.

\begin{figure}
\caption{Atomic s orbitals on the vertices of an icosahedron.}
\label{fig:ico}
\centerline{\epsfysize 3in
\epsfbox{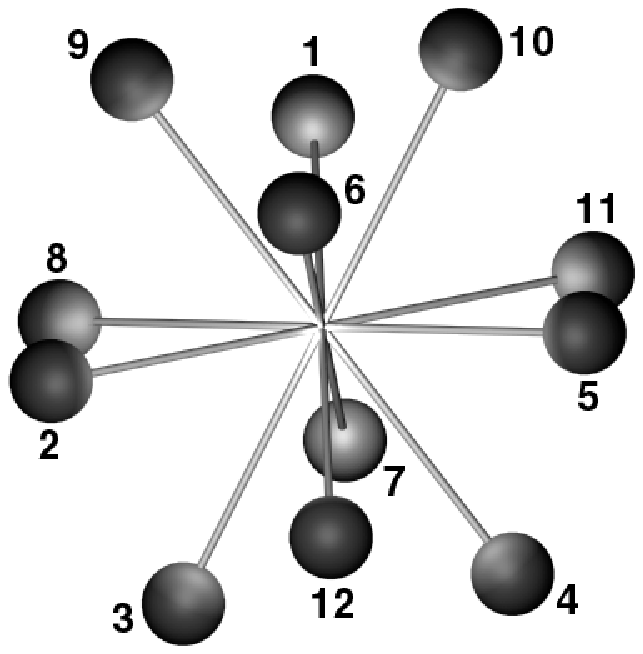}}
\end{figure}

If, on the other hand, we work with molecular $h$ orbitals, the two operators
$e^{(6)}_2$ and $e^{(6)}_3$ (together with $e_0,e_1$ and $e_2$) represent
the first-order Coulomb interaction between the pairs $(i,j)$ of electrons.
To have an image of an $h$ orbital, we follow Plakhutin\cite{Plak},
who constructed icosahedral $g$ orbitals by superposing atomic s orbitals on
the 20 vertices of a dodecahedron.  For $h$ orbitals, we can reduce the basis
to six pairs $\frac1{\sqrt2}(\Psi_n+\Psi_{13-n})$ of identical atomic s
orbitals $\Psi$ on opposite ends ($n$ and $13-n$) of the 6 five-fold axes of
an icosahedron.  This is shown in Fig.~\ref{fig:ico}.  There is only one
complication: the 6 pairs can be combined to produce the icosahedral scalar
$\sum_n\Psi_n$.  The five remaining combinations have to be orthogonalized to
this scalar to produce an $h$ state.

The details of how this is done are not important because our interest lies
in the permutations of the 6 axes among themselves. These operations comprise 
the symmetric group $S_6$.  Among these operations are those that describe 
rotations of the icosahedron as a whole, so $S_6 \supset I$.  But there are 
many operations that twist the icosahedron out of shape and are {\em 
non-feasible}.  The scalar $\sum_n\Psi_n$ is untouched; that is, it belongs
to the IR [6] of $S_6$, in the notation of Littlewood\cite{Little}.  However,
the components of the $h$ state undergo transformations that span the IRs
$(10)^-$ of O(5) and [51] of its subgroup $S_6$\cite{Gard}.
Branching rules for some of the IRs of O(5) occurring in $h^N$ are set out
in Table~\ref{tb:brO5}.

The fact that $I$ is isomorphic to the alternating group
$A_5$\cite{Burrow}, a subgroup of $S_5$ and thus of $S_6$ too, allows us
to include it in the sequence
\begin{equation}
O(5) \supset S_6 \supset S_5 \supset I \ , \label{chainS6}
\end{equation}
which can be considered an alternative to the more familiar scheme
\begin{equation}
SO(5)\supset SO(3)\supset I \ . \label{chainO3}
\end{equation}
The IRs of $S_5$ contained in a particular IR [$\lambda$] of $S_6$ can be
found by interpreting [$\lambda$] as a Young tableau and then removing from 
it a single cell in all possible ways that leave an acceptable shape, with 
the proviso that the automorphism of $S_6$ that produces the interchanges
\begin{equation}
[51] \leftrightarrow [2^3] \ , \ [41^2] \leftrightarrow [31^3] \ , \
[3^2] \leftrightarrow [21^4] \label{auto}
\end{equation}
be carried out first, if [$\lambda$] appears in~(\ref{auto}).  As an example
of the alternative schemes~(\ref{chainS6}) and~(\ref{chainO3}), we express
the linear combination
\begin{equation}
p\{\stackrel +2\! \!-\! \!\stackrel -2\}+q\{\stackrel +1\! \!-\! \!
\stackrel -1\}+r\{\stackrel +0 \stackrel -0\}+q\{-\! \! \stackrel +1
\stackrel -1\}+p\{-\! \!\stackrel +2 \stackrel -2\}
\end{equation}
of Slater determinants for the four states
\begin{eqnarray}
&&|\,{\rm d^2 (20)\,^1D}H\rangle \ ,\ |\,{\rm d^2 (20)\,^1G}H\rangle
\label{std2} \\
&&|\,h^2 (20)^+[51][2^21]^1H\rangle\ ,\ |\,h^2(20)^+[42][32]^1H\rangle
\label{sth2}
\end{eqnarray}
by the respective specifications
\begin{eqnarray}
(p,q,r)&=&\case1{\sqrt{14}}(-2,-1,2) \ , \ \case1{\sqrt{70}}(1,4,6) \ , \\
&&\case1{2\sqrt5}(-1,1,4) \ , \ \case12(1,1,0) \ .
\end{eqnarray}

In our analysis of $g$ orbitals\cite{gorbit}, we introduced the
kaleidoscope operator ${\cal K}$ which rotates the SO(4) weight space by
$\frac{\pi}2$ and which interchanges the $T_1$ and $T_2$ states.  The SO(5)
weight space is very similar to that for SO(4), and we define ${\cal K}$
in the present case by its action on the components $h^\dag_m$ of the
orbital creation tensor $\bbox{h}^\dag$:
\begin{equation}
\begin{array}{lclc}
{\cal K}h^\dag_2{\cal K}^{-1} = h^\dag_1 &,&
{\cal K}h^\dag_{-2}{\cal K}^{-1} = -h^\dag_{-1} &, \\
{\cal K}h^\dag_1{\cal K}^{-1} = h^\dag_{-2} &,&
{\cal K}h^\dag_{-1}{\cal K}^{-1} = -h^\dag_2 &, \\
{\cal K}h^\dag_0{\cal K}^{-1} = h^\dag_0 &.&& \\
\end{array}
\label{Ktrans}
\end{equation}
The five weights $(-2\le m\le2)$ are again rotated by $\frac{\pi}2$, as can 
be seen by inspecting the labels $m$ of the root vectors\cite{OT}. The phases 
in Eq.~(\ref{Ktrans}) have been chosen, first, to preserve the anticommutation
relations for fermions when each $h^\dag_m$ is replaced by $\widetilde h_m \,
(= (-1)^mh_{-m}$); and, second, to give the usual phases for the time-reversal 
operator ${\cal T}$ when we interpret it as ${\cal K}^2$. As in the 
$g$-orbital case, the ${\cal K}$ transformation
changes $m$ to $3m$(mod~5), and, for many-electron states, $M_L$ to
$3M_L$(mod~5).  It again interchanges the states $T_1$ and $T_2$, and we can
now see that there must be as many $T_1$ as $T_2$ states for any $h^N$, a 
result that is not obvious when the SO(3) basis is used.  As for the
states~(\ref{sth2}), ${\cal K}$ sends the first into itself and reverses the 
sign of the second; while the Russell-Saunders states~(\ref{std2}) merely
become mixed.

The icosahedral scalars that appear in the Hamiltonian fall into four
categories.  They belong to $[6][5]A$, $[2^3][5]A$, $[3^2][1^5]A$ or
$[1^6][1^5]A$.  The first two
are invariant under the ${\cal K}$ transformation, while the last two change
sign.  The ${\cal K}$ operation belongs to the class $(1^24)$ and (14) of
$S_6$ and $S_5$ (in the notation of Littlewood\cite{Little}), but it does not
belong to $I$.  The characters of $S_6$ or $S_5$ indicate whether the $A$
component (for which $M_L\equiv 0$(mod~5)) is invariant or changes sign.
We can now use $S_6$ and $S_5$ to recast the three Coulomb operators
$e_2,e^{(6)}_2$ and $e^{(6)}_3$.  The first two belong to the IR $(22)^+$ of
O(5), which, from Table~\ref{tb:brO5}, can be combined into operators
belonging to $[2^3][5]A$ and $[3^2][1^5]A$.
They are given by $\frac17(e_2+\frac{15}2e^{(6)}_2)$ and $\frac17(e_2-3
e^{(6)}_2)$ respectively.
%The actual forms are given
%in the third and fourth columns of Table~\ref{tb:coul}.
The $S_5$ descriptions account for their non-vanishing matrix elements often 
taking complementary
block forms.  The exception is the $T_{1,2}$ states; they belong to the
self-conjugate IR $[31^2]$ of $S_5$, which allows both the scalar [5] and
the pseudo-scalar $[1^5]$ to be non-vanishing at the same time.  The operator
$e^{(6)}_3$ belongs to (40)$^+$[6][5], and is therefore diagonal in the $S_6$
basis.  This operator $e^{(6)}_3$, as well as $e_0, e_1$ and $e_2+\frac{15}2
e^{(6)}_2$, are all invariant under ${\cal K}$; hence we would have three
pairs of $T_{1,2}$ degenerate levels if they were the only operators in the
Hamiltonian.  The presence of the remaining operator $e_2-3e^{(6)}_2$ breaks
this symmetry, but incompletely; leaving us a single $T_{1,2}$ degeneracy as
found by Oliva\cite{Oliva}.
%the entries in the third, fourth and fifth columns of Table~\ref{tb:coul}
%are identical for the last $^2T_1$ and $^2T_2$ levels reproduces the
%single $T_{1,2}$ degeneracy found by Oliva\cite{Oliva}.

It sometimes happens that the $(S_6,S_5)$ classification of a state coincides
with that provided by SO$_L(3)$. In $h^3$, for example, the $^4$P term can be
labelled by $[31^3][31^2]T_1$ (see the first column of Table~\ref{tb:coul}).
An unexpected double labelling occurs for $^2$F and $^2$H terms, whose
$T_2$ components are $[321][31^2]$ and $[41^2][31^2]$ respectively.
To prove this, we first combine the two equations (with $M_L=0$ specified)
\begin{equation}
{\cal K}|{\rm d}^2\,^3{\rm F}\,0\rangle=-|{\rm d}^2\,^3{\rm P}\,0\rangle \ ,
\ {\cal K}|{\rm d}^2\,^3{\rm P}\,0\rangle=|{\rm d}^2\,^3{\rm F}\,0\rangle
\end{equation}
with the invariance of $(\bbox{h}^\dag \bbox{h})^{[51]H}_{\ \ \ \ 0}$ and 
the non-vanishing\cite{Pool} of the icosahedral Clebsch-Gordan coefficient
$(T_10,H0|T_20)$ to show that
\begin{equation}
([41^2]T_1+[51]H\,|\,[41^2]T_2) = 0 \ , \label{3j0}
\end{equation}
using the notation of Racah\cite{Racah} for the isoscalar factor.  It
remains to make the correspondence
\begin{equation}
h^{\dag [51]}\,|\,h^2[41^2]^3T_1\rangle \leftrightarrow
{\rm d^\dag\,|\,d^2\,^3P\rangle} \label{correspond}
\end{equation}
and note that the only state on the left that can contain $^2T_2$ is [321],
since the only other possibility that $[51]\times[41^2]$ can give rise to is
$[41^2]$ (see\cite{Wyb}), and this is excluded by~(\ref{3j0}).  On the right
of~(\ref{correspond}), however, no $^2$H can be produced from angular momenta
of 2 and 1; and so the missing term $[41^2]^2T_2$ is purely $^2$H, and
$[321]^2T_2$ can only belong to the remaining source, namely $^2$F.

The $L$-purity of the $^2T_2$ states of $h^3$ leads to an explanation for
Eqs.~(\ref{t3deg1}) and~(\ref{t3deg2}). The branching~(\ref{chainS6}) for the
IR $(30)^+$ of O(5) yields the labels $[2^3][5]A$ and $[1^6][1^5]A$ for the
two combinations $t_3-10t^{(6)}_3$ and $t_3+18t^{(6)}_3$ given in
Table~\ref{tb:t3}.  Now
\begin{equation}
\langle {\rm F}[321]\,|\,t_3+18t^{(6)}_3\,|\,{\rm H[41^2]}\rangle = 0
\label{t30}
\end{equation}
because $[1^6]\times[41^2]$ does not contain [321].  Also, $\langle{\rm F}
|t_3|{\rm H}\rangle=0$ for the SO(3) scalar $t_3$.  So both parts of the
operator in~(\ref{t30}) yield zero, and so must the combination
\begin{equation}
C = \case9{14} (t_3-10t^{(6)}_3)-\case5{14}(t_3+18^{(6)}_3) \ ,
\end{equation}
which becomes $t_3$ under the ${\cal K}$ transformation.  Thus
\begin{eqnarray}
0&=&\langle{\rm ^2F}T_2\,|\,C\,|\,{\rm ^2H}T_2\rangle=\langle{\rm ^2F}T_2\,|\,
{\cal K}^{-1}{\cal K}C{\cal K}^{-1}{\cal K}\,|\,{\rm ^2H}T_2\rangle\nonumber\\
&=&\langle[321][311]^2T_1\,|\,t_3\,|\,[411][311]^2T_1\rangle \nonumber \\
&=&\left(\alpha\langle{\rm^2P}T_1|+\beta\langle{\rm^2H}T_1|\right)\,t_3\,
\left(\beta|{\rm ^2P}T_1\rangle-\alpha|{\rm ^2H}T_1\rangle\right)\nonumber\\
&=&\alpha\beta\left(\langle{\rm ^2P}\,|\,t_3\,|\,{\rm ^2P}\rangle-
\langle{\rm ^2H}\,|\,t_3\,|\,{\rm ^2H}\rangle\right) \ .
\end{eqnarray}
Neither $\alpha$ nor $\beta$ is zero (being in fact $\sqrt{2/7}$ and
$\sqrt{5/7}$), so Eq.~(\ref{t3deg1}) is recovered.  A similar analysis
for the $G$ states coming from $^2$F and $^2$G yields Eq.~(\ref{t3deg2}).

The success of this analysis suggests that it could be repeated to obtain
Eq.~(\ref{couldeg}).  However, the analog of Eq.~(\ref{t30}) cannot
come directly from $S_6$ because the required null matrix element, although
appearing in Table~\ref{tb:coul}, is not a consequence of a selection rule
for the IRs of $S_6$.  However, $S_6$ immediately yields Eq.~(\ref{t30}) 
for $e^{(6)}_3$, indicating that all we need to do is
prove this equation for either of the two operators prefaced by $a$ and $b$
in Eq.~(\ref{e23}).  If we pick the second and separate both this operator
and the two states $^2$F and $^2$H into spin-up and spin-down parts
(distinguished by the symbols $A$ and $B$\cite{Juddspin}), we are ultimately 
led to calculate a matrix element of the form
\begin{equation}
\langle ({\rm P}_A{\rm d}_B){\rm F}\,|\,(\bbox{A}^{(4)}\bbox{B}^{(4)}
)^{(6)}\,|\,({\rm F}_A{\rm d}_B){\rm H}\rangle \ .
\end{equation}
This is proportional to the vanishing 9-$j$ symbol
\begin{equation}
\left\{\begin{array}{ccc} 1&2&3\\ 4&4&6\\ 3&2&5\end{array}\right\} = 0 \ ,
\end{equation}
and so justifies our again following the procedure starting with the 
analog of Eq.~(\ref{t30}).

What the above analysis shows is that insight into the properties of a free
atom, which are based on SO(3) symmetry, can be gained by using the intrinsic
icosahedral symmetry of SO(3) and, more significantly, the spatial
distortions associated with the group $S_6$ that permutes the axes of the
icosahedron.  We are reluctant to assert that any aspect of the theory we
have presented can be thought of as being caused by another part, since the
theory forms an interlocking whole.  Nevertheless, it is remarkable that
some long-standing puzzles in the classic theory of atomic structure should
receive explanations when $S_6$ is brought into play.

% now the references. delete or change fake bibitem. delete next three
%   lines and directly read in your .bbl file if you use bibtex.

% \begin{table}
% \caption{}
% \label{}
% \end{tabular}
% \end{table}

%****************************************************************
% LaTeX: Branching Rules O(5) to S_6                 Nov '98
%****************************************************************
%
%\arraycolsep= 0.05cm
\renewcommand{\arraystretch}{1.3}
%
%\input newcommand.tex
%\begin{document}
%
\begin{table}
\caption{Branching rules for some IRs $W$ of O(5) into IRs [$\lambda$] of
$S_6$. The signs attached to the IRs of SO(5) indicate the proper (+) or 
improper ($-$) nature of the rotations.}
\label{tb:brO5}
\[ \begin{array}{clcl} \hline \hline
W& [\lambda] &W& [\lambda] \\ \hline
(00)^+&[6] &(00)^- &[1^6] \\
(10)^+&[21^4] &(10)^- &[51] \\
(11)^+&[41^2] &(11)^- &[31^3] \\
(20)^+&[42][51] &(20)^- &[21^4][2^21^2] \\
(21)^+&[2^21^2][31^3][321] &(21)^- &[321][41^2][42] \\
(22)^+&[2^3][321][3^2][42] &(22)^-&[2^21^2][2^3][321][3^2]\\ \hline \hline
\end{array} \]
\end{table}
%
%\end{document}

\newpage
\vspace*{1ex}
\newpage
%****************************************************************
% LaTeX: Coulomb matrix elements for h^3 config      Nov '98
%****************************************************************
%
%\arraycolsep= 0.05cm
\renewcommand{\arraystretch}{1.3}
%
%\input newcommand.tex
%\begin{document}
%
%\vspace{25ex}
\begin{table}
\caption{Some Coulomb matrix elements for the configuration $h^3$.}
\label{tb:coul}
\[ \begin{array}{crccc} \hline \hline
\mbox{Term}& \multicolumn{1}{c}{\mbox{IRs of}} & (22)^+[222][5]A & 
(22)^+[33][1^5]A & (40)^+[6][5]A \\
\mbox{mixtures}& \multicolumn{1}{c}{\mbox{scheme~(\ref{chainS6})}} &\frac17 
\langle e_2+\frac{15}2e^{(6)}_2\rangle & \frac17 \langle
e_2-3e^{(6)}_2\rangle & \langle e^{(6)}_3 \rangle  \\ \hline
^4{\rm F} & (11)^-[31^3][41]^4G\,&-\frac92&0&0\\
^4{\rm P} & (11)^-[31^3][31^2]^4T_1\,&3&3&0\\
^4{\rm F} & (11)^-[31^3][31^2]^4T_2\,&3&-3&0\\
\begin{array}{c} ^2{\rm D}\\ ^2{\rm D,G,H}\\ ^2{\rm D,G,H}\\ 
^2{\rm D,G,H}\end{array}&
\begin{array}{r}(10)^-[51][2^21]^2H\\ (21)^-[321][2^21]^2H\\ 
(21)^-[321][32]^2H\\ (21)^-[42][32]^2H\end{array}&
\left[ \begin{array}{cccc} 0&3\sqrt6&0&0\\
3\sqrt6&-\frac32&0&0\\ 0&0&-\frac{39}{10}&3\sqrt6 \\ 0&0&3\sqrt6&-\frac32
\end{array}\right] &
\left[ \begin{array}{cccc}0&0&\frac{6\sqrt6}5&\frac{18}5 \\ 0&0&\frac95&
\frac{2\sqrt6}5\\ \frac{6\sqrt6}5&\frac95&0&0\\ \frac{18}5&\frac{2\sqrt6}5&0&0 
\end{array}\right] &
\begin{array}{c} 0\\5\\5\\ -\frac{20}3\end{array} \\
\begin{array}{c} ^2{\rm F,G}\\ ^2{\rm F,G}\end{array}&
\begin{array}{r}(21)^-[41^2][21^3]^2G\\ (21)^-[42][41]^2G\end{array}&
\left[\begin{array}{cc}\frac92&0 \\ 0&-\frac12\end{array}\right] 
&\left[ \begin{array}{cc}0&\sqrt{15} \\ \sqrt{15}&0
\end{array}\right] & \begin{array}{c} -2 \\ -\frac{20}3\end{array} \\
\begin{array}{c} ^2{\rm P,H}\\ ^2{\rm P,H}\end{array}&
\begin{array}{r}(21)^-[321][31^2]^2T_1\\ (21)^-[41^2][31^2]^2T_1
\end{array} &\left[ \begin{array}{cc} \frac92&0 \\
0&-3 \end{array} \right] & \left[ \begin{array}{cc} -3&0 \\ 0&0 \end{array}
\right] & \begin{array}{c} 5 \\ -2 \end{array} \\
\begin{array}{c} ^2{\rm F}\\ ^2{\rm H}\end{array}&
\begin{array}{r} (21)^-[321][31^2]^2T_2\\ (21)^-[41^2][31^2]^2T_2
\end{array} &\left[ \begin{array}{cc} \frac92&0 \\
0&-3 \end{array} \right] & \left[ \begin{array}{cc} 3&0 \\ 0&0 \end{array}
\right] & \begin{array}{c} 5 \\ -2 \end{array} \\ \hline \hline
\end{array} \]
\end{table}
\begin{table}
\caption{Some matrix elements for the operators $t_3$ and 
$t^{(6)}_3$ in $h^3$.}
\label{tb:t3}
\[ \begin{array}{rcc} \hline \hline
\multicolumn{1}{c}{\mbox{IRs of}} & (30)^+[222][5]A & (30)^+[1^6][1^5]A \\
\multicolumn{1}{c}{\mbox{scheme~(\ref{chainS6})}}& \frac17 \langle 
t_3-10t^{(6)}_3\rangle & \frac17 \langle t_3+18t^{(6)}_3\rangle \\ \hline
\begin{array}{r}(21)^-[41^2][21^3]^2G\\ (21)^-[42][41]^2G\end{array}&\left[
\begin{array}{cc}-2&0 \\ 0&-2
\end{array}\right] &\left[ \begin{array}{cc}0&0 \\ 0&0 \end{array}\right] \\
\begin{array}{r}(21)^-[321][31^2]^2T_1\\ 
(21)^-[41^2][31^2]^2T_1\end{array} &
\left[ \begin{array}{cc} -\frac13&0 \\ 0&\frac43 \end{array} \right] & \left[
\begin{array}{cc} 3&0 \\ 0&0 \end{array}\right] \\
\begin{array}{r} (21)^-[321][31^2]^2T_2\\ 
(21)^-[41^2][31^2]^2T_2\end{array} & \left[
\begin{array}{cc} -\frac13&0 \\ 0&\frac43 \end{array} \right] & \left[
\begin{array}{cc}-3&0 \\ 0&0 \end{array}\right] \\ \hline \hline
\end{array} \]
\end{table}
%
%\end{document}

\end{document}